\documentclass[11pt]{article}
\input epsf
\begin{document}
\baselineskip 100pt
\renewcommand{\baselinestretch}{1}
\renewcommand{\arraystretch}{0.5}
{\large
\parskip.2in
\newcommand{\be}{\begin{equation}}
\newcommand{\ee}{\end{equation}}
\newcommand{\br}{\bar}
\newcommand{\fr}{\frac}
\newcommand{\lm}{\lambda}
\newcommand{\ra}{\rightarrow}
\newcommand{\al}{\alpha}
\newcommand{\bt}{\beta}
\newcommand{\pr}{\partial}
\newcommand{\hs}{\hspace{5mm}}
\newcommand{\up}{\upsilon}
\newcommand{\dg}{\dagger}
\newcommand{\vphi}{\vec{\phi}}
\newcommand{\ve}{\varepsilon}
\newcommand{\acc}{\\[3mm]}
\newcommand{\dl}{\delta}
\newcommand{\pa}{\partial}
\newcommand{\tht}{\theta}
\newcommand{\ie}{{\it ie }}
\newcommand{\vv}{{\bf V}}
\newcommand{\I}{{\it I}}
\newcommand{\Ic}{{\cal I}}
\newcommand{\N}{{\cal N}}
\newcommand{\cmod}[1]{ \vert #1 \vert ^2 }
\newcommand{\mod}[1]{ \vert #1 \vert }
\newcommand{\nhat}{\mbox{\boldmath$\hat n$}}
\newcommand{\C}{{\rm C}}
\newcommand{\news}{\setcounter{equation}{0}}
\def\U {{\rlap{\kern 1.2mm \vrule height 7pt depth 0pt} \rm 1}}

\nopagebreak[3]

\title{\hbox{\hspace{3mm}{\Large{\bf Spherically Symmetric  Solutions of
the $SU(N)$ Skyrme Models}}
\raisebox{16mm}{\hspace{-40mm}{\large DTP99/17 UKC99/20}\hs\hs \hs\hs
}}}

\author{
T. Ioannidou\thanks{e-mail address:
T.Ioannidou@ukc.ac.uk},\,
\\
Institute of Mathematics, University of Kent at Canterbury, \\
Canterbury CT2 7NF, UK
\acc
\acc
B. Piette\thanks{e-mail address: B.M.A.G.Piette@durham.ac.uk} \,
and
W. J. Zakrzewski\thanks{e-mail address:W.J.Zakrzewski@durham.ac.uk},
\\
Department of Mathematical Sciences,University of Durham, \\
Durham DH1 3LE, UK\\} \date{}

\maketitle
\begin{abstract}

Recently we have presented in \cite{IPZ} an ansatz which allows us to
construct skyrmion fields from the harmonic maps of $S\sp2$ to
$CP\sp{N-1}$.
In this paper  we examine this construction in detail and use it
to  construct, in an explicit form, new static spherically symmetric
solutions of the $SU(N)$ Skyrme models.
We also discuss some properties of these solutions.

\end{abstract}

\section{Introduction}

The Skyrme model presents an opportunity to understand nuclear physics as
a low energy limit of quantum chromodynamics (QCD).
The model was initially proposed as a
theory of strong interactions of hadrons
\cite{Skyr1}, but recently, it was shown to be the low energy limit of QCD
in the large $N_c$ limit \cite{Wit}.
Since then further work has suggested that topologically nontrivial
solutions of this model, known as
skyrmions, can be identified with classical ground states of light
nuclei.
However, a thorough understanding of the structure and dynamics of
multi-skyrmion configurations is required before a more qualitative
assessment of the validity of this application of the model can be made.

The $SU(N)$ Skyrme model involves fields which take values in $SU(N)$;
\ie are described by $SU(N)$ valued functions of $\vec{x}$ and $t$.
Its static solutions correspond to field configurations describing
multi-skyrmions. 
In this paper new solutions have been obtained
for fields whose energy density is spherically symmetric.

Multi-skyrmions are stationary points (maxima or saddle points) of the
static energy functional, which is given in topological charge units by
\be
E=\fr{1}{12\pi^2}\int_{R^3}\left\{-\fr{1}{2}\mbox{tr}\left(\pr_iU\,
U^{-1}\right)^2-\fr{1}{16}\mbox{tr}\left[\pr_iU\,
U^{-1},\;\pr_j U\, U^{-1}\right]^2\right\}d^3\vec{x},
\label{gene}
\ee
where $U(\vec{x})\in SU(N)$.

In this case multi-skyrmions are solutions of the equation
\be
\pr_i\left(\pr_iU\, U^{-1}-\fr{1}{4}\,[\pr_jU\, U^{-1},[\pr_j U\, U^{-1},
\pr_iU\, U^{-1}]]\right)=0.
\label{geq}
\ee
We have, for simplicity, set the mass terms to zero.
This has been done for convenience, since the conventional mass terms
introduce only small changes and, as we will see later, affect only
the profile functions.
Therefore, all our discussion can be easily generalised to 
include such mass terms.

Finiteness of the energy functional requires that $U(\vec{ x})$ approaches
a constant matrix at spatial infinity, which can be chosen to be the
identity matrix by a global $SU(N)$ transformation.
So, without any loss of generality, we can impose
the following boundary condition on $U$: $U \ra {\it
I}$ as $|\vec{ x}|\ra \infty$.

Since  $U \ra {\it I}$ as $|\vec{ x}|\ra \infty$ is a mapping from
$S^3\ra SU(N)$, it can be classified by the third homotopy
group $\pi_3(SU(N))\equiv Z$ or, equivalently, by the integer valued winding
number
\be
B=\fr{1}{24\pi^2}\int_{R^3} \ve_{ijk}\,\mbox{tr}\left(\pr_i U\,
U^{-1}\pr_j U\, U^{-1}
\pr_k U\, U^{-1}\right)d^3\vec{x},\label{bar}
\ee
which is a topological invariant.
This winding number classifies the solitonic sectors in the model,
and as Skyrme has argued \cite{Skyr1},  $B(U)$ may be
identified with the baryon number of the field configuration.

Up to now most of the studies involving the Skyrme model have concentrated 
on the $SU(2)$ version of the model and its embeddings into $SU(N)$.
The simplest nontrivial classical {\it solution} involves a single
skyrmion ($B=1$) and has already been discussed by Skyrme \cite{Skyr1}.
The energy density of this solution is radially symmetric
 and, as a result, using the so-called
hedgehog ansatz one can reduce (\ref{geq}) to an ordinary differential
equation, which  then has to be solved numerically.
Many solutions with $B > 1$ of the $SU(2)$ model have also been computed
numerically and, in all cases, the solutions are very symmetrical (cf.
Battye et al. \cite{BS} and references therein).
However, since the model is not integrable, with few exceptions, explicit
solutions (even) for spherical symmetric $SU(N)$ skyrmions are not known.

The first example of a {\it non-embedded} solution for a higher
group was the $SO(3)$ soliton,  corresponding to a bound system of
two skyrmions, which  was found by Balachandran et al. \cite{BBLRS}.
Another solution, with a large $SU(3)$ strangeness content,
was found by Kopeliovich et al. \cite{Kop}.
However, all  other known multi-skyrmion configurations seem to be
the embeddings of the solutions of the $SU(2)$ model.

Recently, we  have showed in \cite{IPZ} how to construct low energy
states of the $SU(N)$ model by using $CP^{N-1}$ harmonic maps.
Our discussion involved only one projector.
In this paper, we extend our method to more projectors.
We show that, for the $SU(N)$
model, when we take $N-1$ projectors  which lead to spherically symmetric
energy densities, the full equations of the model
separate and the problem
of finding exact solutions is reduced to having to solve $N-1$
coupled nonlinear ODE's for $N-1$ profile functions.
This way we obtain a whole family of new spherical symmetric
multi-baryon solutions of the $SU(N)$ models.
Our solutions include the $SU(3)$ dibaryon configuration of
Balachandran et al. \cite{BBLRS} and the non-topological $SU(3)$ four
baryon configuration of \cite{IPZ}.

\section{Harmonic Maps}

In \cite{IPZ} we generalised the $SU(2)$ ansatz of Houghton et al.
\cite{HMS} to $SU(N)$.
This generalisation involved re-writing the expression
of Houghton et al. as a projector from $S\sp2$ to $CP\sp{N-1}$.
It gave us a new way of interpreting old results
and  of deriving expressions for the low energy $SU(N)$ field configurations
which are {\it not} simple embeddings of $SU(2)$ fields.
In particular, the energy distributions exhibit  very different
symmetries from those of the embeddings.
The method also gave us a new solution of the $SU(3)$ model,
which  lies in the topologically trivial sector of the model ({\it
ie} it has zero baryon number) and so, obviously, is not stable.

The method of \cite{IPZ} can be generalised further, to involve more
projectors.
In fact, we can exploit here some ideas taken from the theory of harmonic maps
of $S\sp2\rightarrow CP\sp{N-1}$ \cite{Zak, DinZak};
since they play an important role in our construction.

Recall (cf. \cite{DinZak}) that in $N$-dimensional space there
is a ``natural" set of projectors: $S\sp2\rightarrow CP\sp{N-1}$ maps,
which are constructed as follows:

Write each projector $P$  as
\be
P(V)=\fr{V \otimes V^\dg}{|V|^2},
\label{for}
\ee
where $V$ is a $N$-component complex vector of  two variables
$\xi$ and $\bar{\xi}$ which  locally parametrise $S\sp2$.
In terms of the more familiar $\theta$ and $\varphi$,  they are
given by $\xi=\tan(\theta/ 2)\,e\sp{i\varphi}$.
The first projector is obtained by taking $V=f(\xi)$, {\ie} an analytic
vector of $\xi$; while the other projectors are obtained
from the original $V$ by differentiation and Gramm-Schmidt
orthogonalisation.
If we define an operator $P_+$  by its action on any vector $v \in \C^N$
\cite{DinZak} as
\be
P_+ v=\pr_\xi v- v\,\fr{v^\dg \,\pr_\xi v}{|v|^2},
\ee
then the further vectors $P^k_+ v$ can be defined by induction:
$P^k_+ v=P_{+}(P^{k-1}_+ v)$.

Therefore, in general, we can consider projectors $P_k$ of the
form (\ref{for}) corresponding to the family of vectors
$V\equiv V_k=P\sp{k}_+f$
(for $f=f(\xi)$) as
\be
P_k=P(P^k_+ f), \hs \hs k=0,\dots,N-1,
\label{maps}
\ee
where, due to the orthogonality of the projectors, we have 
$\sum_{k=0}^{N-1}P_k=1$.

The orthogonality properties of our projectors follow
from the following properties of vectors $P^k_+ f$ 
which hold when $f$ is holomorphic:
\begin{eqnarray}
\label{bbb}
&&(P^k_+ f)^\dg \,P^l_+ f=0, \hs \hs \hs k\neq l,\acc
&&\pr_{\bar{\xi}}\left(P^k_+ f\right)=-P^{k-1}_+ f \fr{|P^k_+
f|^2}{|P^{k-1}_+ f|^2},
\hs \hs
\pr_{\xi}\left(\fr{P^{k-1}_+ f}{|P^{k-1}_+ f|^2}\right)=\fr{P^k_+
f}{|P^{k-1}_+f|^2}.
\label{aaa}
\end{eqnarray}
Note that, for $SU(N)$,  the last projector $P_{N-1}$ in the sequence
corresponds to an anti-analytic vector; ({\ie} the
components of $V_{N-1}=P^{N-1}_+f$, up to an irrelevant
overall factor which cancels in the projector, are functions of only  $\bar{\xi}$).

Our new $SU(N)$ generalisation of \cite{IPZ} involves the introduction of
$N-1$ projectors, \ie
\begin{eqnarray}
U&=&\exp\{ig_0
(P_0-\fr{\I}{N})+ig_1(P_1-\fr{\I}{N})-\dots+ig_{N-2}(P_{N-2}-\fr{\I}{N})\}
\nonumber\\
&=&e^{-ig_0/N}(\I+A_0P_0)\,\,e^{-ig_1/N}(\I+A_1P_1)\,\dots\,
\,e^{-ig_{N-2}/N}(\I+A_{N-1}P_{N-2}),
\label{SUN}
\end{eqnarray}
where $g_k=g_k(r)$, for $k=0, \dots, N-2$, are the profile functions and  
$A_k=e^{ig_k}-1$.
Note that the projector $P_{N-1}$ is not included in the above formula
since it is the linear combination of the others.
[Our previous ansatz given in \cite{IPZ} corresponds to putting all the profile
functions, but the first one, equal to zero.]

The spherically symmetric maps into $CP^{N-1}$ are given by
\begin{eqnarray}
f = (f_0, \,f_1, \dots, \, f_{N-1})^t, \hs \mbox{where} \hs f_k = \xi^k
\sqrt{C_{k+1}^{N-1}},
\label{vec1}
\end{eqnarray}
where $C_{k+1}^{N-1}$ denote the binomial coefficients.
Furthermore, as we prove in the appendix, the modulus of the corresponding
vector $P_+^k f$ for $f$ of the above form is
\be
\vert P\sp{k}_+f\vert\sp2=\alpha(1+\vert \xi\vert\sp2)\sp{N-2k-1},
\ee
where $\alpha$ depends on $N$ and $k$.


\section{Constructing the Skyrmion Solutions}

In this section we construct a family of exact spherical symmetric
solutions of the $SU(N)$ Skyrme models.
In fact, we show that for each $SU(N)$ model the Skyrme field
involving $N-1$ projectors leads to an exact solution involving $N-1$
profile functions.

\subsection{Skyrme Equations}
The Skyrme equations (\ref{geq}), when re-written in spherical
coordinates, take the form:
\begin{eqnarray}
\pa_r\!\left[r\sp2R_r+{1\over 4}\left(A_{\theta r \theta}+{1\over
\sin\sp2 \theta}A_{\varphi r\varphi}\right)\right]
\!\!+\!\!{1\over \sin\theta}\pa_{\theta}\!\left[\sin\theta\left\{R_\theta
+{1\over 4}\left(A_{r\theta r}+{1\over r\sp2\sin\sp2\theta}A_{\varphi
\theta\varphi}\right)\right\}\right]&&\nonumber \\
+{1\over \sin\sp2\theta} \pa_\varphi\!\left[R_\varphi+{1\over 4}
\left(A_{r\varphi r}+{1\over
r\sp2}A_{\theta\varphi\theta}\right)\right]=0,&&
\label{eqqq}
\end{eqnarray}
where $R_i=U\sp{-1}U_i$ and
$A_{\alpha\beta\gamma}\equiv [R_\alpha,\,[R_\beta,\,R_\gamma]].$

It is easy to see, using (\ref{SUN}), that
\begin{equation}
\label{rr}
R_r=i\sum_{j=0}\sp{N-2}\dot g_j\left(P_j-{\I\over
N}\right),
\end{equation}
where $\dot g_j(r)$ denotes the derivative of $g_j(r)$ with respect to its
argument; and that, in terms of the holomorphic variables $\xi$ and $\bar \xi$,
\begin{eqnarray}
R_\xi&=&e^{(-i\sum_{k=0}\sp{N-2}g_kP_k)}\,
\pr_\xi\!\left[e^{(i\sum_{i=0}\sp{N-2}g_iP_i)}\right]\nonumber\\
&=&\left[1+\sum_{k=0}\sp{N-2}(e\sp{-ig_k}-1)P_k\right]
\left[\sum_{l=0}\sp{N-2}(e\sp{ig_l}-1)P_{l\xi}\right]\nonumber\\
&=&
\sum_{i=1}\sp{N-1}\left[e\sp{i(g_i-g_{i-1})}-1\right]{V_i\,V\sp{\dagger}_{i-1}
\over \vert V_{i-1}\vert\sp2},
\label{rxi}
\end{eqnarray}
where the last line follows from  the identity
$e^{(-i\sum_{k=0}\sp{N-2}g_kP_k)}=1+\sum_{k=0}\sp{N-2}(e\sp{-ig_k}-1)P_k$.
Here,  $g_{N-1}=0$ and  $R_{\bar\xi}=-(R_\xi)\sp{\dagger}.$

Next we note that
\be
\pa_{\theta}=\fr{1+|\xi|^2}{2\sqrt{|\xi|^2}}\left(\xi
\,\pa_\xi\,+\bar\xi\,\pa_{\bar\xi}\right),\hs \hs
\pa_{\varphi}=i\left(\xi \,\pa_\xi\,-\bar\xi\,\pa_{\bar\xi}\right),
\label{mig}
\ee
and re-write all the terms in (\ref{eqqq}) in terms of
$R_\xi$, $R_{\bar \xi}$ and $R_r$ and their commutators, {\it ie}
\begin{eqnarray}
&&\hspace{-13mm}\pr_{\tht} \left(\sin \tht\, R_{\tht}\right)+\fr{1}{\sin
\tht}\pr_{\phi}
R_{\phi}=(1+|\xi|^2)\sqrt{|\xi|^2}\left((R_\xi)_{\bar{\xi}}
+(R_{\bar{\xi}})_{\xi}\right),\\
&&\label{En}\hspace{-13mm}A_{\theta r \theta}+{1\over
\sin\sp2\theta}A_{\varphi
r\varphi}
=\fr{(1+|\xi|^2)^2}{2}\left\{[R_{\bar\xi},[R_r,R_\xi]]
+[R_\xi,[R_r,R_{\bar\xi}]]\right\},\acc
&&\hspace{-13mm}\sin \tht \,\pr_{\tht} \left(\sin \tht\,A_{r\theta
r}\right)+\pr_{\phi}(A_{r\varphi r})=2|\xi|^2
\left([R_r,[R_{\xi},R_r]]_{\bar{\xi}}
+[R_r,[R_{\bar{\xi}},R_r]]_{\xi}\right),\\
&&\hspace{-13mm}\pr_{\tht}\!\!\left(\!\fr{A_{\varphi \theta\varphi}}{\sin
\tht}\!\right)\!\!+\!\!\fr{1}{\sin
\tht}\pr_{\phi}\!\!\left(A_{\theta\varphi\theta}\right)\!=\!
\fr{(1+|\xi|^2)\,\sqrt{|\xi|^2}}{2}\!\!\left[\pr_{\bar{\xi}}\left((1+|\xi|^2)^2
[R_\xi,[R_\xi,R_{\bar\xi}]]\right)\!\!-\!\!\pr_{\xi}\!\left((1+|\xi|^2)^2
[R_{\bar\xi},[R_{\xi},R_{\bar\xi}]]\right)\!\right].\nonumber\\
\end{eqnarray}

Thus  equation (\ref{eqqq}), when re-written in the holomorphic variables,
becomes
\begin{eqnarray}
&&\pa_r\left[r\sp2R_r+
\fr{(1+|\xi|^2)^2}{8}\left([R_{\bar\xi},[R_r,R_\xi]]
+[R_\xi,[R_r,R_{\bar\xi}]]\right)\right]
+\fr{(1+|\xi|^2)^2}{2}\left((R_{\bar{\xi}})_\xi+
(R_{\xi})_{\bar{\xi}}\right)+\nonumber\\[1.5mm]
&&\fr{(1+\vert\xi\vert\sp2)\sp3}{8r^2}\left(\xi\,
[R_\xi,[R_\xi,R_{\bar\xi}]]-\bar\xi\,
[R_{\bar\xi},[R_\xi,R_{\bar\xi}]]\right)
+{(1+\vert \xi\vert\sp2)\sp4\over
16r\sp2}\left([R_\xi,[R_{\xi},R_{\bar\xi}]]_{\bar\xi}-[R_
{\bar\xi},[R_\xi,R_{\bar\xi}]]_{\xi}\right)\nonumber\\[1.5mm]
&&+{(1+\vert\xi\vert\sp2)\sp2\over 8}
\left([R_{r},[R_{\bar\xi},R_{r}]]_\xi
+[R_r,[R_\xi,R_{r}]]_{\bar\xi}\right)=0.
\label{Eqq}
\end{eqnarray}

Using (\ref{SUN}) we observe that
\begin{eqnarray}
&&[R_\xi,\,R_{\bar{\xi}}]=2\,P_0{\vert V_1\vert\sp2\over
\vert V_0\vert\sp2}\left(1-\cos(g_1-g_0)\right)\,-2\,P_{N-1}{\vert V_{N-1}
\vert\sp2\over \vert
V_{N-2}\vert\sp2}\left(1-\cos(g_{N-2})\right)+\nonumber\\
&&\hspace{20mm}\,2\,\sum_{i=1}\sp{N-2}\,P_i\,
\left[{\vert V_{i+1}\vert\sp2\over \vert
V_i\vert\sp2}\left(1-\cos(g_{i+1}-g_i)\right)
-{\vert V_i\vert\sp2\over \vert V_{i-1}\vert\sp2}
\left(1-\cos(g_i-g_{i-1})\right)\right]\!\!,\label{ncomm}\hs\acc
&&[R_{\xi},[R_\xi,\,R_{\bar{\xi}}]]=
\sum_{i=1}\sp{N-1}
\left(\mu_i{\vert V_{i+2}\vert\sp2\over \vert V_{i+1}\vert\sp2}
+\nu_i{\vert V_{i+1}\vert\sp2\over \vert V_{i}\vert\sp2}
+\rho_i{\vert V_{i}\vert\sp2\over \vert V_{i-1}\vert\sp2}\right)
{V_iV\sp{\dagger}_{i-1} \over \vert V_{i-1}\vert\sp2},\acc
&&[R_r,[R_{\bar\xi},R_r]]=\sum_{i=1}
\sp{N-1}s_i\,{V_{i-1}V\sp{\dagger}_i\over\vert V_{i-1}\vert\sp2},
\end{eqnarray}
where $\mu$, $\nu$ and $\rho$ are functions of $g_k(r)$, only; while
$s_i$ are functions of $g_k(r)$ and their derivatives.

Since $V_k=P_+^k f$, one can show that ${\vert V_{i}\vert\sp2\over
\vert V_{i-1}\vert\sp2}\propto (1+\vert \xi\vert\sp2)^{-2}$; while
$\pa_\xi\,(1+\vert \xi\vert\sp2)^{-2}=-2\bar\xi
(1+\vert\xi\vert\sp2)^{-3}$ and thus, the derivative terms involving
$[R_\xi,[R_\xi,R_{\bar\xi}]]$ in (\ref{Eqq}) cancel 
leaving us with 
derivatives of ${V_iV\sp{\dagger}_{i-1} \over \vert V_{i-1}\vert\sp2}$ --
which are proportional to
$\sum_{i=1}\sp{N-1}(1+\vert\xi\vert\sp2)^{-2}(P_i-P_{i-1})\,h_i$,
where $h_i$ involve functions of $g_k(r)$ (due to (\ref{aaa})), multiplied
by terms of the form  ${\vert V_{i}\vert\sp2\over\vert
V_{i-1}\vert\sp2}$.
So the factors $(1+\vert\xi\vert\sp2)^{-4}$ in ({\ref{Eqq}) cancel
-- leaving us with a sum of differences of two successive
projectors multiplied by functions dependent only on $r$.
 
Following the above argument and  using the properties of $R_r$, {\it
etc } one can show that the terms $[R_r,[R_{\xi},R_r]]_{\bar\xi}$ in
(\ref{Eqq}), are proportional to $\sum_{i=1}\sp{N-1}S_i\,
(1+\vert\xi\vert\sp2)^{-2}(P_i-P_{i-1})$, where $S_i$ are functions
of $g_k(r)$ and their derivatives -- leaving us, once again,  with a sum
of differences of two successive projectors multiplied by functions
dependent only on $r$.

Finally, the contribution of the terms
$(1+|\xi|^2)\,R_{\xi\bar\xi}$ is given by $\sum_{i=1}\sp{N-1}
(P_i-P_{i-1}) H_i,$ where $H_i$ are only functions of $g_k$; while
the commutators in (\ref{En}) are equal to a sum of
projectors mutlipied by $(1+\vert\xi\vert\sp2)^{-2}$, which 
cancel out in (\ref{Eqq}).
In addition,  $\pa_r( r\sp2 R_r)=i
\sum_{i=0}\sp{N-2}\left(P_i-{\I\over N}\right)(2r\dot g_i+r\sp2\ddot
g_i).$

We note that, for our choice of the vectors $V_k$, all the dependence
on $\xi$ and $\bar\xi$ in (\ref{Eqq}) resides only in the
projectors (the rest of it cancels out).
The terms involving $\pa_r(r\sp2 R_r)$ give us expressions involving
${1\over N}-P_i$ while all the other terms give us expressions involving
$P_i-P_{i-1}$.
Although $N$ projectors arise in (\ref{Eqq}),
the projector $P_{N-1}$ can be re-expressed in terms
of the previous ones -- giving $N-1$ factors involving
the harmonic maps $P_i-{1\over N}$ (for $i=0,... N-2$).
To satisfy (\ref{Eqq}) the  coefficients of such factors
have to vanish leaving us with $N-1$ equations for the $N-1$ profile
functions $g_i$.
Hence, if these equations have solutions then they correspond to exact
solutions of the $SU(N)$ Skyrme models.
Notice that (\ref{vec1}) implies that these solutions have a covariant axial 
symmetry, {\it ie} any rotation by an angle $\alpha$ around the $z$-axis
is  equivalent to the gauge transformation $U \rightarrow A^{\dagger}U A$
where $A = \mbox{diag}(1,e^{i\alpha}, e^{2 i \alpha}, ..
e^{(N-1)i\alpha})$.
On the other hand, as will be shown below, the energy density for these 
solution is radially symmetric.   

The $N-1$ equations for the profile functions can be obtained either from
(\ref{Eqq}) -- which is a hard task; or from the variation of the energy
(\ref{gene}) -- using (\ref{SUN}) and integrating out $\xi$ and $\bar\xi$
variables.
Clearly, the two methods give the same equations.

Let us stress that our procedure hinges on having $N-1$ profile functions
and on the very special form of our vectors $V_k$. 
Had we taken other vectors $V_k$, we would have got some $\xi$
and $\bar\xi$ dependence outside the projectors; while had we taken less
than $N-1$ profile functions and projectors we would have got too many
equations for our functions.
It is only in the case of $N-1$ projectors that we get the right number
of equations.

\subsection{Energy Dependence on Profile Functions}

The energy (\ref{gene}), when written in the holomorphic variables, becomes 
\begin{equation}
\label{energy}
E\!=\!-\fr{i}{12\pi^2}\!\!\int\! r^2 dr\,d\xi d\bar{\xi}
\,\mbox{tr}\left(\fr{1}{(1+\vert \xi\vert\sp2)\sp2}R\sp2_r+
\fr{1}{r\sp2} \vert R_\xi\vert\sp2
+\fr{1}{4r^2}[R_r,R_\xi][R_r, R_{\bar\xi}]-\fr{(1+\vert
\xi\vert\sp2)\sp2}{16r\sp4}[R_{\bar\xi},R_\xi]\sp2\right).
\label{pen}
\end{equation} 

Using (\ref{rr}) and  (\ref{rxi}) we find that
\begin{eqnarray}
&&\mbox{tr}\,R_r\sp2={1\over N}\left(\sum_{i=0}\sp{N-2}\dot
g_i\right)\sp2\,-\,\sum_{i=0}\sp{N-2}\dot g_i\sp2,\\
&&\mbox{tr}\vert R_\xi\vert\sp2\,=\,-2\sum_{i=1}\sp{N-1}
B_i,\\
&&\mbox{tr}[R_r,R_\xi][R_r,R_{\bar\xi}]=-2\sum_{k=1}\sp{N-1}
B_k(\dot g_k-\dot g_{k-1})\sp2,\\
&&\mbox{tr}[R_{\bar\xi},R_\xi]\sp2=4\left(B_1\sp2 +\sum_{i=1}\sp{N-2}
(B_i-B_{i+1})\sp2+B_{N-1}\sp2\right),
\end{eqnarray}
where $B_i={\vert V_i\vert\sp2\over \vert
V_{i-1}\vert\sp2}(1-\cos(g_i-g_{i-1}))$
and  $[R_{\bar\xi},R_\xi]=2\sum_{l=1}\sp{N-1}(P_{l-1}-P_l)B_l$.

Since ${\vert V_k\vert\sp2\over
\vert V_{k-1}\vert\sp2}=k(N-k)(1+\vert \xi\vert\sp2)^{-2}$ (see appendix)
 all
terms in (\ref{pen}) have a factor
$(1+\vert \xi\vert\sp2)^{-2}$ and the integration over $\xi$ and $\bar\xi$
is a topological constant, {\it ie} $i\int d\xi d\bar{\xi} (1+\vert
\xi\vert\sp2)^{-2}\,=\,2\pi$.
Thus we get
\begin{eqnarray}
E\!\!\!\!&=&\!\!\!\fr{1}{6\pi}\int\!\! r\sp2 dr\{-{1\over
N}\left(\sum_{i=0}\sp{N-2}
\dot g_i\right)\sp2+\sum_{i=0}\sp{N-2}\dot g_i\sp2 +{1\over
2r^2}\sum_{k=1}
\sp{N-1}\left(\dot g_k-\dot g_{k-1}\right)\sp2D_k
+{2\over r\sp2} \sum_{k=1}\sp{N-1}D_k\nonumber\\
&&\hspace{17mm}+{1\over 4r\sp4}
\left(D_1\sp2+\sum_{k=1}\sp{N-2}(D_k-D_{k+1})\sp2+D_{N-1}\sp2\right)\},
\end{eqnarray}
where $D_k=k(N-k)(1-\cos(g_k-g_{k-1}))$.

Let us, for simplicity, take $F_k=g_k-g_{k+1}$, ($k=0,..N-2)$
with $F_{N-2}=g_{N-2}$.
Then, the variation of  the integrand  of the energy $\tilde E$ with
respect to the functions $\dot F_l$ (for  $l=0,..N-2$) is
\be
{\partial \tilde E\over \partial \dot F_l}=\left[-{2(l+1)\over N}\sum_{i=
0}
\sp{N-2}(i+1)\dot F_i+2\sum_{k=0}\sp{l}\left(\sum_{i=k}\sp{N-2}\dot F_i\right)
+\fr{1}{r^2}\dot F_l D_{l+1}\right]r\sp2,
\ee
where $D_k=k(N-k)\left(1-\cos F_{k-1}\right)$.

Therefore, the equations of motion for the functions $F_i$, and
thus for the profile functions, are
\begin{eqnarray}
&&-{2(l+1)\over N}\sum_{i=0}\sp{N-2}(i+1)\ddot F_i+2\sum_{k=0}\sp{l}
\sum_{i=k}\sp{N-2}\ddot F_i+\fr{1}{r^2}\ddot
F_l(l+1)(N\!-\!l\!-\!1)(1-\cos
F_l)+\nonumber\\
\label{profs}
&&{1\over 2r\sp2}\dot F_l\sp2(l\!+\!1)(N\!-\!l\!-\!1)\sin F_l\!+\!{2\over
r}\left(
\!-{2(l\!+\!1)\over N}\sum_{i=0}
\sp{N-2}(i+1)\dot F_i\!+\!2\sum_{k=0}\sp{l}\left(\sum_{i=k}\sp{N-2}\dot
F_i\right)\right)\nonumber\\
&&-{2\over r\sp2}\,(l+1)(N-l-1)\,\sin F_l-
{1\over r\sp4}\,(l+1)\sp2(N-l-1)\sp2\left(1-\cos F_l\right)\sin
F_l\nonumber\\
&&+{1\over 2r\sp4}(l\!+\!1)(N\!-\!l\!-\!1)\sin F_l\left[l(N\!-\!l)(1-\cos
F_{l-1})+(l\!+\!2)(N\!-\!l\!-\!2)(1-\cos F_{l+1})\right]=0.\nonumber\\
\end{eqnarray}
These equations can  be solved numerically by imposing the appropriate 
boundary conditions on the profile functions.
To do this we have to specialise to a particular model, \ie for specific 
$N$ and diagonalise the terms involving the second derivatives.
The simplest cases: $N=2$, $N=3$ and $N=4$ involve $1$, $2$ or $3$
functions and will be discussed in the next sections.

\section{Topological Properties and Symmetries}

Before we discuss special cases, let us first investigate the
general topological properties of our fields.

The topological charge (\ref{bar}), which in many applications of the Skyrme
model is identified with the baryon number, is given by
\be
B=\fr{1}{8\pi^2}\int dr\, d\xi d\bar{\xi}
\,\mbox{tr}\, \left(R_r\,[R_{\bar\xi},R_\xi]\right),
\ee
when written in the complex coordinates.

Due to (\ref{ncomm}) and (\ref{rr}) the terms involving $\dot g_i/N$ in $R_r$
after taking the trace cancel and the expression for the
baryon number simplifies to
\begin{eqnarray}
B&=&-\fr{i}{4\pi^2}\int dr\, d\xi d\bar{\xi}
\,\sum_{i=0}\sp{N-2}\, \dot F_i\,(1-\cos F_i){\vert V_{i+1}\vert\sp2\over
\vert V_i\vert\sp2}\nonumber\\
&=&\fr{1}{2\pi}\int dr \sum_{i=0}\sp{N-2}\dot F_i\,(1-\cos
F_i)(i+1)(N-i-1)\nonumber \\
&=& \fr{1}{2\pi}\sum_{i=0}\sp{N-2}(i+1)(N-i-1)\,\left(F_i-\sin
F_i\right)_{r=0}\sp{r=\infty}.
\label{BofF}
\end{eqnarray}
As $g_i(\infty)=0$ (required for the finiteness
of the energy) the only contributions to the topological charge
come from $F_i(0)$.

The $N-1$ equations for the profile functions and their differences
given in (\ref{profs})  have many symmetries.
These symmetries can be used to derive special skyrmion solutions
which involve a smaller number of profile functions and projectors.

The main symmetry of our expressions, are the  independent interchanges
\be
F_k\,\leftrightarrow\,F_{N-k-2},\qquad \mbox{for}\hs k=0,\cdots, N-2.
\ee
This symmetry follows from the fact the terms $D_k=k(N-k)(1-\cos
F_{k-1})$ which appear in the energy are symmetric under the interchange:
 $D_k\leftrightarrow D_{N-k}$ when $F_{k-1}\leftrightarrow F_{N-k-1}$.
In addition, all the other terms in the energy also exhibit this
symmetry since they are combinations of $F_i$ and their derivatives.

\section{Spherical Skyrmions}

The simplest case corresponds to the $SU(2)$ spherically symmetric skyrmion.
This is the solution which was found thirty years ago by
Skyrme and is usually presented in terms of the well-known hedgehog ansatz.

\subsection{$SU(3)$ Skyrme Model}

In this case $N=3$ and we have two functions: $F_0$ and $F_1$.
The energy density ${\cal E}$, such that $E = (6\pi)^{-1}\int {\cal E}
r^2 dr$, is given by 
\begin{eqnarray}
{\cal E}\!\!\!\!&=&\!\!\!\!
{2\over 3}(\dot F_0\sp2+\dot F_1\sp2+\dot F_0\dot F_1) +
\fr{1}{r^2}\left[(\dot F_0\sp2+4)(1-\cos F_0)+
                 (\dot F_1\sp2+4)(1-\cos F_1)\right]+\nonumber\\
\!\!\!\!&& 
\qquad\fr{2}{r^4}\left[(1-\cos F_0)^2-(1-\cos F_0)\,(1-\cos F_1)+
                 (1-\cos F_1)^2\right],\nonumber\\
\end{eqnarray}
and the equations  for the profile functions are
\begin{eqnarray}
&&\hspace{-10mm}
\ddot F_0 \left(\!1\!+\!\fr{3}{2r^2}(1\!-\!\cos F_0)\!\right)+
\fr{\ddot F_1}{2}\!+\!\fr{2\dot F_0\!+\!\dot F_1}{r}+
\fr{3\sin F_0}{4r^2}\left[\dot F_0^2\!-4\!-\!\fr{4(1\!-\!\cos F_0)}{r^2}\!+\!
\fr{2(1\!-\!\cos F_1)}{r^2}\right]\!\!=\!0,\nonumber\\
&&\hspace{-10mm}
\ddot F_1 \left(\!1\!+\!\fr{3}{2r^2}(1\!-\!\cos F_1)\!\right)+
\fr{\ddot F_0}{2}+ \fr{2\dot F_1\!+\!\dot F_0}{r}\!+\!
\fr{3\sin F_1}{4r^2}\left[\dot F_1^2\!-4\!-\!\fr{4(1\!-\!\cos F_1)}{r^2}\!+\!
\fr{2(1\!-\!\cos F_0)}{r^2}\right]\!\!=\!0.\nonumber\\
\label{SU3}
\end{eqnarray}
These equations can be solved numerically when the right boundary
conditions have been imposed.

However, by letting $F_0=F_1=F$ ({\it ie} using the symmetry)
they reduce to
\begin{equation}
\ddot F \left(1+\fr{1-\cos
F}{r^2}\right)+\fr{2\dot F}{r}+
\fr{\sin F}{2r^2}\left[\dot F^2-4-\fr{2(1-\cos F)}{r^2}\right]=0.
\label{max}
\end{equation}
This equation coincides with the equation for the profile function
of a single $SU(2)$ skyrmion. 
Here we note that as $F_0(0)=F_1(0)=2\pi$ the topological charge 
of our solution is four.
Thus the energy of this configuration, which corresponds to four
skyrmions is $E_{B=4}=4\times\,1.232$, {\it ie} is exactly four times the
energy of one skyrmion. 
We see that we have a static solution corresponding to four
non-interacting skyrmions, placed on top of each other in such a way
that their energy density is spherically symmetric.

In addition, there is a further symmetry which allows us to set $F_0=-F_1=G$.
In this case the equations reduce to
\begin{equation}
\ddot G \left(\fr{1}{2}+\fr{3}{2r^2}(1-\cos
G)\right)+\fr{\dot G}{r}+
\fr{3\sin G}{4r^2}\left[\dot G^2-4-\fr{2(1-\cos G)}{r^2}\right]=0.
\end{equation}
Let us note that, since $F_0=g_0-g_1$ and $F_1=g_1$,
this case corresponds to $g_0=0$ and thus, the field (\ref{SUN}) involves only
one projector, namely $P_1$.
This solution is the topologically trivial solution discussed
in \cite{IPZ} and its energy is $3.861$.

Finally, Balachandran et al. skyrmion solution can be obtained from
(\ref{SU3}) by imposing the following boundary conditions: $g_0(0)=2\pi$,
$g_1(0)=0$ and $g_0(\infty)=0$, $g_1(\infty)=0$; its energy is 
$E_{B=2}=2.3764$.

\subsection{$SU(4)$ Skyrme Model}

In this case the energy density becomes
\begin{eqnarray}
&&{\cal E}=
{1\over 4}\left(3\dot F_0\sp2+4\dot F_1\sp2+3\dot F_2^2+4\dot F_0\dot F_1+
4\dot F_1\dot F_2+2\dot F_0\dot F_2\right)+\nonumber\\
&&\hspace{8mm}{1\over 2 r^2}\left[3(\dot F_0\sp2+4)(1\!-\!\cos F_0)\!+
\!4(\dot F_1\sp2+4)(1\!-\!\cos F_1)\!+
\!3(\dot F_2^2+4)(1\!-\!\cos F_2)\right]\nonumber\\
&& \hspace{8mm}+\fr{1}{2r^4}\,\{9\,(1-\cos F_0)^2+16\,(1-\cos F_1)^2+
9\,(1-\cos F_2)^2\nonumber\\
&&\hspace{8mm}
- 12 (1-\cos F_0) (1-\cos F_1)-12(1-\cos F_1)(1-\cos F_2)\},
\end{eqnarray}
while the equations for $F_0$, $F_1$ and $F_2$ are more complicated:
\begin{eqnarray}
&&\hspace{-20mm}
\ddot F_0\left(\!1\!\!+\!\!\fr{2(1-\cos F_0)}{r^2}\!\right)\!+
\!\fr{2 \ddot F_1\!+\!\ddot F_2}{3}\!+\!\fr{3\dot F_0\!+
      \!4\dot F_1\!+\!2\dot F_2}{3r}\!+
\!\fr{\sin F_0}{r^2}
\left[\dot F_0^2\!-\!4\!-\!\fr{6(1\!\!-\!\!\cos F_0)}{r^2}\!+
      \!\fr{4(1\!\!-\!\!\cos F_1)}{r^2}\right]\!\!=\!0,\nonumber\\
&&\hspace{-20mm}
\ddot F_1\!\left(\!1\!\!+\!\!\fr{2(1\!\!-\!\!\cos F_1)}{r^2}\!\right)\!\!+
\!\fr{\ddot F_0\!+\!\ddot F_2}{2}\!+
\!\fr{2\dot F_1\!+\!\dot F_0\!+\!\dot F_2}{r}\!+
\!\fr{\sin F_1}{r^2}\!
\left[\!\dot F_1^2\!-\!4\!-\!\fr{8(1\!\!-\!\!\cos F_1)}{r^2}\!+
      \!\fr{3(1\!\!-\!\!\cos F_0)}{r^2}\!+
      \!\fr{3(1\!\!-\!\!\cos F_2)}{r^2}\!\right]\!\!=\!0,\nonumber\\
&&\hspace{-20mm}
\ddot F_2\left(\!1\!\!+\!\!\fr{2(1\!\!-\!\!\cos F_2)}{r^2}\!\right)\!+
\!\fr{\ddot F_0\!+\!2\ddot F_1}{3}\!+
\!\fr{6\dot F_2\!+\!2\dot F_0\!+\!4\dot F_1}{3r}\!+
\!\fr{\sin F_2}{r^2}
\left[\dot F_2^2\!-\!4\!-\!\fr{6(1\!\!-\!\!\cos F_2)}{r^2}\!+
      \!\fr{4(1\!\!-\!\!\cos F_1)}{r^2}\right]\!\!=\!0.\nonumber\\
\label{SU4eq}
\end{eqnarray}
These equations have the previously mentioned symmetry $F_k
\leftrightarrow F_{N-k-2}$ which allows us to set $F_0=F_2$ by keeping
$F_1$ arbitrary.

In addition, letting $F_0=F_1=F_2=F$ the above system reduces to equation
(\ref{max}) and therefore, the configuration which consists of ten skyrmions
(as $B={3F_0(0)+4F_1(0)+3F_2(0)\over 2\pi}=10$)
is $E_{B=10}=10\times\,1.232$, \ie is exactly ten times the energy of one 
skyrmion. Once again we have a solution describing
 many skyrmions, which are 
non-interacting and whose energy density is spherically symmetric.

In addition, letting $F_0=-F_2=G$ we spot that when
$F_1=0$, we have a solution of the form
\be
\ddot G\left(1+\fr{3(1-\cos G)}{r^2}\right)+\fr{2\dot G}{r}
+\fr{3\sin G}{2r^2}\left[\dot G^2-4-\fr{6(1-\cos
G)}{r^2}\right]=0,
\ee
which corresponds to a non-topological solution, \ie its baryon number
is zero; however the corresponding configuration consists of three
skyrmions and three anti-skyrmions.
[Recall, that the profile functions are  $g_0=0$ and $g_1=g_2$, \ie the
field (\ref{SUN}) involves  only one projector of rank two -- namely
$P_1+P_2$.]
	
In general, however, the solutions depend on more functions. 
We can always  assume that the functions $F_i$ go to zero  at infinity, so
the topological
charge of a solution is determined, using (\ref{BofF}), by the boundary
value of each $F_i$ at the origin. 
When each of these values is positive the solution is a mixture of
skyrmions. 
When some $F_i$'s take positive and some $F_i$'s take negative values at
the origin the solution corresponds  to a mixture of skyrmions and 
anti-skyrmions. 
This is very similar to what happens in the two-dimensional 
Grasmannian sigma model \cite{Zak}. 

We have solved numerically equations (\ref{SU4eq}) taking all combinations,
modulo the exchange of $F_0$ and $F_2$, of $0$, $2\pi$ and $-2\pi$ for 
the value of $F_i$ at the origin. The results are summarised in the
Table below.

\begin{center}
\begin{tabular}{r|r|r|r|l|l|l} 
\vbox to 5mm {}$F_0(0)$ & $F_1(0)$ & $F_2(0)$ & 
B & Energy & E/baryon & SU(2) En \\
&&&&&&\\
\hline
\hline
\vbox to 4mm{}$2\pi$ & 0 & 0 & 3 & 3.518 & 1.173 & 3.429 \\
\vbox to 4mm{}0 & $2\pi$ & 0 & 4 & 4.788 & 1.197 & 4.464 \\
\vbox to 4mm{}$2\pi$ & 0 & $2\pi$ & 6 & 7.22553 & 1.204 & 6.654 \\
\vbox to 4mm{}$2\pi$ & 2$\pi$ & 0 & 7 & 8.45219 & 1.207 & 7.7693 \\
\vbox to 4mm{}$2\pi$ & $2\pi$ & $2\pi$ & 10 & 12.32 & 1.232 & - \\
&&&&&&\\
\hline
\vbox to 4mm{}$2\pi$ & $-2\pi$ & $2\pi$ & 6-4 & 8.852 & 0.8852 & - \\
\vbox to 4mm{}$2\pi$ & $2\pi$ & $-2\pi$ & 7-3 & 9.896 & 0.9896 & - \\
\vbox to 4mm{}$2\pi$ & 0 & $-2\pi$ & 3-3 & 6.63422 & 1.106 & - \\
\vbox to 4mm{}$-2\pi$ & $2\pi$ & 0 & 4-3 & 6.61478 & 0.945 & - \\
\end{tabular}
\end{center}

The first five solutions are bound states of skyrmions with energies 
larger than the energies of the corresponding $SU(2)$ solutions \cite{BS}. 
Moreover, the excitation energy of the first two solutions is very small.
As mentioned above, the energy of the $B=10$ solution is exactly ten times
the  energy of a single skyrmion solution. 
These solutions are all axially symmetric (but their energy
densities are radially symmetric) and thus 
they are all more symmetrical than the corresponding $SU(2)$ solutions.

The last four solutions are bound states of skyrmions and anti-skyrmions. 
Although their energies are very small, we know that these solutions must be 
unstable.
 
\section{Conclusions}

In this paper we have shown how to construct radially symmetric solutions
of the $SU(N)$ Skyrme models. 
In the general case these solutions depend on $N-1$ profile functions
which have to be determined numerically. 
In some cases we can exploit symmetries of our expressions
and reduce the number of functions. 
Thus in the case of $SU(3)$ we can recover the topologically trivial
solution discussed in \cite{IPZ}.

We have not discussed the derived solutions in much detail.
Their properties and their relation to physics deserve
further study and these topics are currently under investigation.

It is worth pointing out that there is a rather close connection
between $SU(N)$ BPS monopoles and skyrmions. Both monopole and skyrmion
fields can be constructed in terms of harmonic maps between Riemann spheres.  
Thus, recently, it has been shown \cite{Sut} that the monopoles fields
can also be constructed using the projector ansatz. 
Its generalisation to multi-projectors and the construction  in
\cite{Sut} provide explicit examples of spherically symmetric $SU(N)$
monopoles with various symmetry breaking patterns.
In the monopole case, the Bogomolny equation is the analogue of our
Skyrme equation and the monopole number corresponds to our baryon number.

\section{ Acknowledgments}

We thank C. Houghton, V. Kopeliovich and P. Sutcliffe for their interest.

\appendix
\section*{Appendix}
\news
\renewcommand{\theequation}{A.\arabic{equation}}

In this appendix we prove equation (11) for $f$ given by (10).
This result can be proved by induction.
Note that, the modulus of the
$N$-dimensional vector $f$ is  $\vert f\vert\sp2=(1+|\xi|\sp2)\sp{N-1}$
and therefore, $\vert P_+f\vert\sp2\vert f\vert\sp2=\vert f\vert \sp2\vert
\partial_+f\vert\sp2 -\vert
f\sp{\dagger}\partial_+f\vert\sp2=(N-1)(1+\vert \xi\vert\sp2)\sp{N-3}\vert f
\vert\sp2$, therefore, $\vert P_+f\vert\sp2=(N-1)(1+\vert
\xi\vert\sp2)\sp{N-3}.$

As
\be
P\sp{k+1}_+f=\pr_\xi P\sp{k}_+f- P\sp{k}_+f{ (P\sp{k}_+f)^\dg\,
\pa_\xi P\sp{k}_+f\over \vert  P\sp{k}_+f\vert\sp2},
\ee
using (\ref{aaa}) we get
\begin{eqnarray}
\vert  P\sp{k+1}_+f\vert \sp2\vert  P\sp{k}_+f\vert\sp2&=&
\vert \pa_\xi P\sp{k}_+f\vert\sp2 \vert  P\sp{k}_+f\vert \sp2-
\vert \pa_\xi \vert  P\sp{k}_+f\vert\sp2\vert\sp2,\acc
\pa_\xi\pa_{\bar{\xi}}\vert  P\sp{k}_+f\vert\sp2&=&\vert\pa_\xi
P\sp{k}_+f\vert \sp2+ ( P\sp{k}_+f)^\dg\,\pa_{\bar{\xi}}\pa_\xi
P\sp{k}_+f,\acc
(P\sp{k}_+f)^\dg\,\pa_\xi\pa_{\bar{\xi}} (P\sp{k}_+f)&=&-(
P\sp{k}_+f)^\dg \, \pa_\xi (P\sp{k-1}_+f)\,{\vert
P\sp{k}_+f\vert\sp2\over
\vert  P\sp{k-1}_+f \vert\sp2}\nonumber\\
&=&-{\vert  P\sp{k}_+f\vert\sp4\over \vert P\sp{k-1}_+f\vert\sp2}.
\end{eqnarray}
Which finally, leads to
\be
\vert  P\sp{k+1}_+f\vert \sp2=\pa_\xi\pa_{\bar{\xi}}\vert  P\sp{k}_+f\vert
\sp2
+{\vert  P\sp{k}_+f\vert \sp4\over \vert  P\sp{k-1}_+f\vert\sp2}
-{\vert \pa_\xi\vert  P\sp{k}_+f\vert\sp2\vert\sp2\over \vert
P\sp{k}_+f\vert\sp2}.
\ee
Therefore if $\vert
P\sp{k}_+f\vert\sp2=\alpha(1+\vert \xi\vert\sp2)\sp{N-2k-1}$
and $\vert P\sp{k-1}_+f\vert\sp2=\beta(1+\vert \xi\vert\sp2)\sp{N-2k+1}$
then
\be
\vert P\sp{k+1}_+f\vert\sp2=\gamma\,(1+\vert \xi\vert\sp2)\sp{N-2k-3},
\ee
where $\gamma=\alpha(N-2k-1)+\alpha\sp2/ \beta$.
To find $\gamma$ we again use induction, recalling that the coefficients
of the two lowest terms in the sequence are 1 and $N-1$, respectively.
Then it is easily seen that
\be
\vert P\sp{k}_+f\vert\sp2 = k!(N-1)(N-2)\cdots (N-k)
(1+\vert \xi\vert\sp2)\sp{N-2k-1},
\ee
in which  the last term in the sequence corresponds to
$k=N-1$.

\end{document}